\documentclass{elsart}
\usepackage{epsfig}
\usepackage{graphicx}
\usepackage{subfigure}

\begin{document} 

\begin{frontmatter}

\title{\bf Highly-anisotropic and strongly-dissipative hydrodynamics with transverse expansion  \thanksref{grant}} 
\thanks[grant]{Supported in part by the Polish Ministry of Science and Higher Education, grants  No. N N202 263438 and N N202 288638.}

\author[ifj]{Radoslaw Ryblewski}
\ead{Radoslaw.Ryblewski@ifj.edu.pl},
\author[ifj,ujk]{Wojciech Florkowski} 
\ead{Wojciech.Florkowski@ifj.edu.pl}

\address[ifj]{The H. Niewodnicza\'nski Institute of Nuclear Physics, Polish Academy of Sciences, ul. Radzikowskiego 152, 31-342 Krak\'ow, Poland}
\address[ujk]{Institute of Physics, Jan Kochanowski University,
ul.~\'Swi\c{e}tokrzyska 15, 25-406~Kielce, Poland}

\begin{abstract} 
A recently formulated framework of highly-anisotropic and strongly-dissipative hydrodynamics (ADHYDRO) is used to describe the evolution of matter created in ultra-relativistic heavy-ion collisions. New developments of the model contain: the inclusion of asymmetric transverse expansion (combined with the longitudinal boost-invariant flow) and comparisons of the model results with the RHIC data, which have become possible after coupling of ADHYDRO with THERMINATOR. Various soft-hadronic observables (the transverse-momentum spectra, the elliptic flow coefficient $v_2$, and the HBT radii) are calculated for different initial conditions characterized by the value of the initial pressure asymmetry. We find that as long as the initial energy density profile is unchanged the calculated observables remain practically the same. This result indicates the insensitivity of the analyzed observables to the initial anisotropy of pressure and suggests that the complete thermalization of the system may be delayed to easily acceptable times of about \mbox{1 fm/c}.
\end{abstract}
\end{frontmatter}
\vspace{-7mm} PACS: 25.75.-q, 25.75.Ld, 24.10.Nz, 25.75.Nq

\section{Introduction}
\label{sect:intro}

The soft-hadronic observables measured in ultra-relativistic heavy-ion collisions at RHIC (transverse-momentum spectra, the elliptic flow coefficient $v_2$, and the HBT radii) are described very well by the perfect-fluid hydrodynamics or by viscous hydrodynamics with a small viscosity to entropy ratio (for recent reviews see, for example, \cite{Florkowski:2010zz} and \cite{Heinz:2009xj}, respectively).  The implicit in this success is the use of a rather early thermalization time that is quite often well below 1 fm~\footnote{We use the natural system of units where $c=k_B=\hbar=1$.}  \cite{Pratt:2008qv,Bozek:2009ty}. The early starting time of hydrodynamics, that is identified with the thermalization time, is justified in a natural way if the quark-gluon plasma is a strongly interacting system \cite{Gyulassy:2004zy,Shuryak:2004cy}. 

On the other hand, there are attempts to describe the RHIC data with the help of models which do not assume very fast thermalization. The success of such models in reproducing the data may indicate that the quark-gluon plasma is not necessarily a strongly interacting system. An example of such a model has been introduced in  Ref. \cite{Broniowski:2008qk}, where the authors assume that the perfect-fluid stage is preceded by free streaming of partons, see also \cite{Sinyukov:2006dw,Gyulassy:2007zz,Qin:2010pf}. Another example is a model discussed in Ref. \cite{Ryblewski:2010tn}, where the initial stage consists of partons with thermalized transverse degrees of freedom only \cite{Bialas:2007gn}.

Recently, we have formulated a framework of highly-anisotropic and strongly-dissipative hydrodynamics (ADHYDRO) \cite{Florkowski:2010cf,Ryblewski:2010bs,Ryblewski:2010ch}. This model interpolates between a  highly-anisotropic  initial state  (where the longitudinal and transverse pressures may be substantially different from each other) and the regime described by the perfect-fluid hydrodynamics. It may be used to analyze the effects of early anisotropic pressure on the subsequent evolution of matter. ADHYDRO may be treated as an important generalization of the models introduced in Refs. \cite{Broniowski:2008qk,Ryblewski:2010tn} because the change from the early highly-anisotropic system to the subsequent locally isotropic stage is described as a continuous process. We note that in \cite{Broniowski:2008qk,Ryblewski:2010tn} a similar change was described with the help of the Landau matching conditions which demand that the energy flux in the transition is conserved but do not demand the continuity of all components of the energy-momentum tensor. 

Our formulation of the model in \cite{Florkowski:2010cf,Ryblewski:2010bs} has been followed by the papers by Martinez and Strickland \cite{Martinez:2010sc,Martinez:2010sd}, where a similar approach has been derived from the kinetic theory with the collision term treated in the relaxation time approximation. In the case of purely longitudinal expansion one may show that the approaches introduced in \cite{Florkowski:2010cf,Ryblewski:2010bs} and \cite{Martinez:2010sc,Martinez:2010sd} lead to the same structure of equations that determine the evolution of pressure asymmetry and entropy density. The only difference may be found in the form of the entropy source. Interestingly, if the pressure asymmetry is small, both approaches are strictly equivalent and they agree with the second-order Israel-Stewart theory. In Ref. \cite{Ryblewski:2010ch} we have shown additionally, that the entropy source terms used in \cite{Florkowski:2010cf,Ryblewski:2010bs} and \cite{Martinez:2010sc,Martinez:2010sd} lead to practically the same results even if the initial pressure asymmetry is very large. 

In this paper, besides the boost-invariant longitudinal motion, the transverse expansion is also taken into account. In such a general case, the explicit connections between ADHYDRO, the underlying kinetic theory, and the Israel-Stewart theory have not been established yet. However, in the initial stage where the effects of pressure asymmetry are most significant, the longitudinal motion dominates the dynamics of the system, and we deal with a situation that is quite similar to that described in Refs.  \cite{Florkowski:2010cf,Martinez:2010sc}. On the other hand, at the stage where the transverse flow becomes substantial, the pressure asymmetries are eliminated and the system's expansion is well described by the perfect-fluid hydrodynamics. Therefore, we consider ADHYDRO as a good candidate for the model which extrapolates well between an initial highly-anisotropic state and a later perfect-fluid phase~\footnote{A natural connection between ADHYDRO and dissipative hydrodynamics may be achieved if the model is constructed explicitly in such a way that it interpolates between an initial highly-anisotropic state and a later dissipative stage described by the Israel-Stewart theory. This work is in progress. }. 

In this paper we use ADHYDRO and show that as long as the initial energy density profile is unchanged the calculated observables remain practically the same. This result indicates the insensitivity of the analyzed observables to the initial anisotropy and suggests that the complete thermalization of the system may be delayed to easily acceptable times of about 1 fm. In this way we find further support to ideas presented in \cite{Broniowski:2008qk,Ryblewski:2010tn}.

The fact that different initial pictures  lead to the same results for hadronic observables means that the freeze-out conditions for different physical scenarios are similar. This suggests the formation of universal flow in heavy-ion collisions, as pointed out first in Refs. \cite{Vredevoogd:2008id,Vredevoogd:2009zu}. On the other hand, different initial conditions may affect other observables, that have not been studied here. In particular, the early pressure anisotropy may have an important impact on the directed flow, as shown recently in \cite{Bozek:2010aj}. The study of this problem requires, however, a generalization of the ADHYDRO code to 3+1 dimensions, which is a work in progress. 

\section{Definition of the model}
\label{sect:modeldef}

The ADHYDRO model is based on the following form of the energy-momentum tensor \cite{Florkowski:2010cf},
\begin{eqnarray}
T^{\mu \nu} &=& \left( \varepsilon  + P_\perp\right) U^{\mu}U^{\nu} - P_\perp \, g^{\mu\nu} - (P_\perp - P_\parallel) V^{\mu}V^{\nu}. 
\label{Tmunudec}
\end{eqnarray}
Here $\varepsilon$, $P_\perp$, and $P_\parallel$ are the energy density, transverse pressure, and longitudinal pressure, respectively. In the case where the two pressures are equal, \mbox{$P_\perp=P_\parallel=P$}, we recover the energy-momentum tensor of the perfect fluid. The four-vector $U^\mu$ in (\ref{Tmunudec}) is the hydrodynamic flow, 
\begin{equation}
U^\mu = \gamma (1, v_x, v_y, v_z), \quad \gamma = (1-v^2)^{-1/2},
\label{Umu}
\end{equation}
and $V^\mu$ defines the longitudinal (beam) direction, 
\begin{equation}
V^\mu = \gamma_z (v_z, 0, 0, 1), \quad \gamma_z = (1-v_z^2)^{-1/2}.
\label{Vmu}
\end{equation}
The four-vectors $U^\mu$ and $V^\mu$ satisfy the following normalization conditions:
\begin{eqnarray}
U^2 = 1, \quad V^2 = -1, \quad U \cdot V = 0.
\label{UVnorm}
\end{eqnarray}
In the local-rest-frame (LRF) of the fluid element we have $U^\mu = (1,0,0,0)$ and \mbox{$V^\mu = (0,0,0,1)$}, and the energy-momentum tensor is reduced to a simple form,

\begin{equation}
T^{\mu \nu}_{\rm LRF} =  \left(
\begin{array}{cccc}
\varepsilon & 0 & 0 & 0 \\
0 & P_\perp & 0 & 0 \\
0 & 0 & P_\perp & 0 \\
0 & 0 & 0 & P_\parallel
\end{array} \right).
\label{Tmunuarray}
\end{equation}
Thus, the formula (\ref{Tmunudec}) allows for different pressures in the longitudinal and transverse directions. 

Besides the energy-momentum tensor (\ref{Tmunudec}), we introduce the entropy flux
\begin{eqnarray}
\sigma^{\mu} &=& \sigma U^{\mu},
\label{smudec}
\end{eqnarray}
where $\sigma$ is the entropy density. We assume that $\varepsilon$ and  $\sigma$ are functions of $P_\perp$ and $P_\parallel$. For massless partons the condition $T^\mu_{\,\,\,\mu}=0$ gives $\varepsilon = 2 P_\perp + P_\parallel$, however, in our actual calculations we use more involved expressions introduced below in Sect. \ref{sect:eos}

The space-time evolution of the system is governed by the equations expressing the energy-momentum conservation and the entropy growth,

\begin{eqnarray}
\partial_\mu T^{\mu \nu} &=& 0, \label{enmomcon} \\
\partial_\mu \sigma^{\mu} &=& \Sigma. \label{engrow}
\end{eqnarray}
The function $\Sigma$ represents the entropy source. The form of $\Sigma$ must be treated as an assumption that defines the dynamics of the anisotropic fluid. It is natural to assume that $\Sigma \geq 0$, with \mbox{$\Sigma = 0$} for \mbox{$P_\perp=P_\parallel$}. In this way, in the case where the two pressures are equal, the structure of the perfect-fluid hydrodynamics is recovered. 

Treating $\Sigma$ as a function of $P_\perp$ and $P_\parallel$, Eqs. (\ref{enmomcon}) and (\ref{engrow}) become a closed system of 5 equations for 5 unknown functions: three components of the fluid velocity, $P_\perp$, and $P_\parallel$. The projections of Eq. (\ref{enmomcon}) on $U_\nu$ and $V_\nu$ yield

\begin{eqnarray}
&& \mathcal{D} \varepsilon = - (\varepsilon + P_{\perp}) \Delta + (P_{\perp}-P_{\parallel}) U_{\nu} V^{\mu}\partial_{\mu}V^{\nu}, \label{enmomconU} \\
&& V^{\mu} \partial_{\mu} P_{\parallel} =  (\varepsilon + P_{\perp}) V_{\nu} \mathcal{D} U^{\nu} + (P_{\perp}-P_{\parallel}) \partial_{\mu} V^{\mu}, \label{enmomconV}
\end{eqnarray}
where we have introduced the operators $\mathcal{D} \equiv U^{\mu} \partial_{\mu} $ and  $\Delta \equiv \partial_{\mu} U^{\mu}$.

\section{Implementation of boost-invariance}
\label{sect:bi}

In this paper we consider boost-invariant systems which expand arbitrarily in the transverse plane. Therefore, we introduce the following parameterizations of the four-vectors $U^\mu$ and $V^\mu$,

\begin{eqnarray}
U^{\mu} &=& (u_0 \cosh\eta , u_x, u_y, u_0 \sinh\eta ),
\label{Uv}
\end{eqnarray}
\begin{eqnarray}
V^{\mu} &=& ( \sinh\eta, 0, 0, \cosh\eta),
\label{Vv}
\end{eqnarray}
where $u_0$, $u_x$, and $u_y$ are components of the flow in the plane $z=0$, satisfying the normalization condition
\begin{equation}
u_0^2 - u_x^2 -u_y^2 = 1,
\label{smallu}
\end{equation}
and $\eta$ is the space-time rapidity 
\begin{eqnarray}
\eta = \frac{1}{2} \ln \frac{t+z}{t-z}. 
\label{eta} 
\end{eqnarray}
Equations (\ref{Uv}) and (\ref{Vv}) satisfy automatically the normalization conditions (\ref{UVnorm}). In the boost-invariant case the derivative $\mathcal{D}$ and the divergence $\Delta$  have the form 
\begin{eqnarray}
\mathcal{D} &=& U^{\mu} \partial_{\mu} = 
u_0 \partial_{\tau} + {\bf{u}}_{\perp} \cdot {\bf{\nabla}}_{\perp}, \\  
\Delta &=& \partial_{\mu} U^{\mu}  = \partial_{\tau} u_0  + 
\frac{u_0}{\tau} + {\bf{\nabla}_{\perp}} \cdot {\bf{u}_{\perp}}, 
\label{help1}
\end{eqnarray}
where ${\bf{u}}_{\perp} = (u_x,u_y)$, ${\bf{\nabla}}_{\perp}=(\partial_x,\partial_y)$, and $\tau$ is the longitudinal proper time
\begin{equation}
\tau = \sqrt{t^2 - z^2}.
\label{tau}
\end{equation} 
Moreover, for the boost-invariant flow we find
\begin{eqnarray}
V^{\mu} \partial_{\mu} &=& \frac{\partial_{\eta}}{\tau}, \quad \partial_{\mu} V^{\mu}  = 0.
\label{help2}
\end{eqnarray}

Equations (\ref{help2}) may be used to check that Eq. (\ref{enmomconV}) is automatically fulfilled for a boost-invariant system. Thus, from the energy-momentum conservation we obtain only three independent equations, the formula (\ref{enmomconU}) which we repeat here for completeness
\begin{eqnarray}
\mathcal{D} \varepsilon &=& - (\varepsilon + P_{\perp}) \Delta + (P_{\perp}-P_{\parallel}) U_{\nu} V^{\mu}\partial_{\mu}V^{\nu}, \label{enmomconUr} 
\end{eqnarray}
and two equations describing transverse dynamics. The latter can be chosen as the linear combinations: $U_{1} \partial_{\mu} T^{\mu 1} +   U_{2} \partial_{\mu} T^{\mu 2} = 0$ and $U_{2} \partial_{\mu} T^{\mu 1} -  U_{1} \partial_{\mu} T^{\mu 2} = 0$, which leads us to the two expressions
\begin{eqnarray}
	& & \mathcal{D} u_{\perp} = - \frac{u_{\perp}}{\varepsilon + P_{\perp}} 
\left[ \frac{{\bf{u}}_{\perp} \cdot {\bf{\nabla}}_{\perp} P_{\perp}}{u_{\perp}^2} 
+ \mathcal{D} P_{\perp} + \frac{u_0}{\tau} (P_{\perp} - P_{\parallel})\right],
	\label{HydEqEuler1}
	 \\
	 & & \mathcal{D} \left( \frac{u_x}{u_y} \right) = \frac{1}{u_y^2 (\varepsilon + P_{\perp})} \left( u_x \partial_y - u_y \partial_x \right)P_{\perp}.
\label{HydEqEuler2}
\end{eqnarray}
Together with the entropy production equation, see Eq. (\ref{engrow}), that can be put in the form
\begin{equation}
\mathcal{D} \sigma + \sigma \Delta = \Sigma,
\label{entrprod}
\end{equation}
we have 4 equations that should determine the dynamics of 2 thermodynamics-like parameters (for example $P_\perp$ and $P_\parallel$) and 2 components of the transverse flow ($u_x$ and $u_y$).

\section{Generalized equation of state}
\label{sect:eos}

The equations of perfect-fluid hydrodynamics form a closed system of equations if they are supplemented with the equation of state that specifies thermodynamic quantities as functions of one parameter, e.g., the system's temperature (for systems with non vanishing baryon number two parameters are necessary). In the similar way, our framework requires that all thermodynamics-like quantities can be expressed as functions of two arbitrarily chosen parameters. Such relations play a role of the generalized equation of state and allow us to close the system of dynamic equations formulated in the previous sections.   

In our previous studies \cite{Florkowski:2010cf,Florkowski:2009sw} we have shown that it is better to switch from $P_\perp$ and $P_\parallel$ to two other thermodynamics-like parameters: the non-equilibrium entropy density  $\sigma$ and the anisotropy parameter $x$. In particular, if we deal with partons described by the anisotropic phase-space distribution function obtained by squeezing/stretching the longitudinal momentum in the Boltzmann distribution by the factor $\sqrt{x}$, we obtain 

\begin{eqnarray}
\varepsilon &=&  \left(\frac{\pi^2 \sigma}{4 g_0} \right)^{4/3} R(x),
\label{epsilon2} 
\end{eqnarray}
\begin{eqnarray}
P_\perp &=&  \left(\frac{\pi^2 \sigma}{4 g_0} \right)^{4/3}
\left[\frac{R(x)}{3} + x R^\prime(x) \right],   
\label{PT2} 
\end{eqnarray}
\begin{eqnarray}
P_\parallel &=&  \left(\frac{\pi^2 \sigma}{4 g_0} \right)^{4/3} 
\left[\frac{R(x)}{3} - 2 x R^\prime(x) \right]. \label{PL2} \\ \nonumber
\end{eqnarray}
The function $R(x)$ is defined by the formula \cite{Florkowski:2009sw} 
\begin{eqnarray}
R(x) = \frac{3\, g_0\, x^{-\frac{1}{3}}}{2 \pi^2} \left[ 1 + \frac{x \arctan\sqrt{x-1}}{\sqrt{x-1}}\right],
\label{Rx}  \\ \nonumber
\end{eqnarray}
and $g_0$ is the degeneracy factor connected with internal quantum numbers of partons. The symbol $R'(x)$ denotes the derivative of $R(x)$ with respect to $x$, for $x=1$ we have $R'(1)=0$ and the two pressures are equal.

We stress that Eqs. (\ref{epsilon2})--(\ref{PL2}) describe the non-equilibrium state defined by the values of the non-equilibrium entropy density $\sigma$ and the anisotropy parameter $x$. However, since in the equilibrium we have:  $\varepsilon_{\rm id} = 3 g_0 T^4/\pi^2$, $P_{\rm id} = g_0 T^4/\pi^2$, and $\sigma_{\rm id} = 4 g_0 T^3/\pi^2$ \cite{Florkowski:2010zz},  Eqs. (\ref{epsilon2})--(\ref{PL2}) may be written equivalently in the form~\footnote{For simplicity, Eqs. (\ref{epsilon2})--(\ref{PL2}) have been introduced for the squeezed/stretched classical (Boltzmann) distribution. A generalization to the squeezed/stretched quantum (Bose-Einstein or Fermi-Dirac) distributions has been given recently in \cite{Ryblewski:2010ch}. This generalization introduces an irrelevant factor. },
\begin{eqnarray}
\varepsilon &=&  \varepsilon_{\rm id}(\sigma) r(x), \label{epsilon2a}  \\ \nonumber \\
P_\perp &=&  P_{\rm id}(\sigma) \left[r(x) + 3 x r^\prime(x) \right],  \label{PT2a}  \\ \nonumber \\
P_\parallel &=&  P_{\rm id}(\sigma) \left[r(x) - 6 x r^\prime(x) \right], \label{PL2a} 
\end{eqnarray}
where $\varepsilon_{\rm id}(\sigma)$ and $P_{\rm id}(\sigma)$ are equilibrium expressions for the energy density and pressure, and $r(x) = \pi^2 R(x)/ (3 g_0)$. One can easily notice that for $x=1$ we have $r(1)=1$ and $r^\prime(1)=0$, hence Eqs. (\ref{epsilon2a})--(\ref{PL2a}) are reduced to the equilibrium expressions. 

As stated above, the structure of Eqs. (\ref{epsilon2})--(\ref{PL2})  and Eqs. (\ref{epsilon2a})--(\ref{PL2a}) follows from the special form of the momentum distribution. This form is very much likely to be realized at the very early stages of the collisions. On the other hand, at later stages of the collisions, the produced matter approaches local equilibrium described by the appropriate equation of state. In other words, as the system becomes more isotropic, i.e., as the parameter $x$ tends to unity, the local properties of the fluid become closer to those characterized by the realistic equation of state. This argument convinced us to use, in the numerical calculations presented below, the following expressions
\begin{eqnarray}
\varepsilon &=&  \varepsilon_{\rm qgp}(\sigma) r(x), \label{epsilon2b}  \\ \nonumber \\
P_\perp &=&  P_{\rm qgp}(\sigma) \left[r(x) + 3 x r^\prime(x) \right], \label{PT2b}   \\ \nonumber \\
P_\parallel &=&  P_{\rm qgp}(\sigma) \left[r(x) - 6 x r^\prime(x) \right], \label{PL2b} 
\end{eqnarray}
where $\varepsilon_{\rm qgp}(\sigma)$ and $P_{\rm qgp}(\sigma) $ characterize the realistic equation of state for vanishing baryon chemical potential, as it has been constructed in Ref. \cite{Chojnacki:2007jc}, and the function $r(x)$ is the same as that appearing in Eqs. (\ref{epsilon2a})--(\ref{PL2a}).

It is necessary to admit that Eqs. (\ref{epsilon2b})--(\ref{PL2b}) have no direct microscopic explanation. The main motivation for this form is that it has two attractive limits. At the very early stage the system consists most likely of massless partons and Eqs. \mbox{(\ref{epsilon2b})--(\ref{PL2b})} approach Eqs. \mbox{(\ref{epsilon2a})--(\ref{PL2a})}, since the realistic equation of state approaches the Stefan-Boltzmann limit (although deviations from this limit are noticeable). On the other hand, when the system becomes isotropic, $x=1$ and  Eqs. \mbox{(\ref{epsilon2b})--(\ref{PL2b})}  are reduced to the equation of state used in standard hydrodynamics. 

\section{Entropy source}

In this paper, following our original formulation of the model in Ref. \cite{Florkowski:2010cf}, we use the following form of the entropy source
\begin{equation}
\Sigma = \frac{(1-\sqrt{x})^2}{\sqrt{x}} \frac{\sigma}{\tau_{\rm eq}}.
\label{ansatz1}
\end{equation} 
The quantity ${\tau_{\rm eq}}$ is a timescale parameter. The form (\ref{ansatz1}) guarantees that $\Sigma \geq 0$ and $\Sigma(\sigma,x=1)=0$. More arguments for such a particular form of $\Sigma$ are given in \cite{Florkowski:2010cf}. They connect $x$ with the ratio of the transverse and longitudinal temperatures. 

In the case of purely longitudinal expansion and for small deviations from equilibrium one can show that Eq. (\ref{ansatz1}) leads to quadratic dependence of the entropy source on the variable $\xi=1-x$ and this dependence is compatible with the Israel-Stewart theory (where the entropy production depends on the viscous stress squared) and with the Martinez-Strickland model \cite{Martinez:2010sc}. For more general situations, the equation (\ref{ansatz1}) should be treated as one of the assumptions defining our model. Other forms of the entropy source have been analyzed in \cite{Ryblewski:2010ch}. All those forms lead to similar numerical results. 

We emphasize that the structure of the entropy source is an external input for the anisotropic hydrodynamics. Especially, in the region where the asymmetries are large and no correspondence to dissipative hydrodynamics can be found. In this context, it is interesting to obtain any hints about $\Sigma$ for large $x$ from the microscopic models of particle production or from the AdS/CFT correspondence. 

In our numerical calculations we adopt the value $\tau_{\rm eq} = 0.25$ fm. This choice is suggested by our previous calculations \cite{Florkowski:2010cf,Ryblewski:2010bs,Ryblewski:2010ch}, where we have shown that it leads to almost complete thermalization of matter at the proper time of about 1 fm (for highly anisotropic initial conditions studied here). Our earlier calculations have shown that reaching the perfect-fluid regime at $\tau \sim 1$ fm is necessary to obtain a good agreement with the data~\footnote{If the free-streaming or transverse-hydrodynamics stage is longer than 1 fm, it is difficult to achieve a good description of data, as discussed in Refs. \cite{Broniowski:2008qk,Ryblewski:2010tn}.}. 

The value $\tau_{\rm eq} = 0.25$ fm corresponds to $\tau^{\rm coll}_{\rm eq} = (2/15) \tau_{\rm eq} \approx 0.03$ fm used in the collision term by Martinez and Strickland \cite{Ryblewski:2010ch,Martinez:2010sc}. Such a small value of $\tau^{\rm coll}_{\rm eq}$ indicates that the plasma is indeed a strongly interacting system. However, our example shows that the average collision time in a highly-anisotropic plasma and the plasma isotropization time may be quite different.

\section{Anisotropy evolution}

If both the generalized equation of state and the entropy production term are defined, one can derive a compact expression for the time evolution of the anisotropy parameter $x$. Of course, this expression should be considered together with other dynamic equations. Nevertheless, its form turns out to be useful in the analysis of the time evolution of the system.

In the case described by Eqs. \mbox{(\ref{epsilon2b})--(\ref{PL2b})}, we find
\begin{eqnarray}
   \mathcal{D} x &=&  \frac{3 x P_{\rm qgp}}{\varepsilon_{\rm qgp}} \left( \frac{3 u_0}{\tau} -\Delta \right) - \left(1 + \frac{P_{\rm qgp}}{\varepsilon_{\rm qgp}}\right) \frac{H(x)}{\tau_{\rm eq}},
   \label{x1}
\end{eqnarray}
where 
\begin{equation}
H(x) =  \frac{r(x)}{r'(x)} \frac{(1-\sqrt{x})^2}{\sqrt{x}}.
\label{H}
\end{equation}
If $3 P_{\rm qgp}=\varepsilon_{\rm qgp}$, which is the case realized by Eqs. \mbox{(\ref{epsilon2a})--(\ref{PL2a})}, Eq. (\ref{x1}) is reduced to the form
\begin{eqnarray}
\mathcal{D} x &=& x \left( \frac{3 u_0}{\tau} -\Delta \right) -\frac{4}{3 \tau_{\rm eq}} H(x).
\label{x2}
\end{eqnarray}
For the purely longitudinal boost-invariant motion, $u_0=1$ and $\Delta = 1/\tau$, and Eq. (\ref{x2}) simplifies to an ordinary differential equation for $x$, 
\begin{equation}
\frac{dx}{d\tau}  =  \frac{2 x}{\tau} - \frac{4 H(x)}{3\tau_{\rm eq}}.
\label{x3}
\end{equation}
We shall come back to the discussion of Eq. (\ref{x3}) in Sect. \ref{centr}.

\section{Initial conditions}

The initial conditions for the evolution are defined by four functions: $\sigma(\tau_0,{\bf x}_\perp)$, $x(\tau_0,{\bf x}_\perp)$, $u_x(\tau_0,{\bf x}_\perp)$, and $u_y(\tau_0,{\bf x}_\perp)$, where $\tau_0$ is the initial proper time. In the numerical calculations we assume \mbox{$\tau_0 =$ 0.25 fm}. We also assume that there is no transverse flow present initially, therefore we set $u_x(\tau_0,{\bf x}_\perp)=0$ and $u_y(\tau_0,{\bf x}_\perp)=0$. 

For the initial  anisotropy $x(\tau_0,{\bf x}_\perp)$, which we take as independent of ${\bf x}_\perp$  and denote simply as $x_0$, we consider three different options: $x_0=100$, $x_0=1$, and $x_0=0.032$. The case $x_0=1$ is, of course, the closest to that described by standard perfect-fluid hydrodynamics. 

The case $x_0=100$ corresponds to the initial situation where the transverse pressure is much larger than the longitudinal pressure. In this case the momentum shape is {\it oblate}; the momentum distribution is stretched in the transverse direction and squeezed in the longitudinal direction to the beam. This type of the initial conditions is considered in the Color Glass Condensate (CGC) approach where the distribution functions in the longitudinal direction are described by the Dirac delta function, $\delta(p_\parallel)$ at $z=0$ \cite{Kovner:1995ja,Bjoraker:2000cf},  see also Ref. \cite{El:2007vg}. The values $x_0 < 1$  correspond to the {\it prolate} momentum shape. This type of the initial conditions have been analyzed, for example, in Refs. \cite{Jas:2007rw,Randrup:2003cw}. 

The value $x_0=0.032$ has been chosen, since $r(100) \approx r(0.032)$, and for fixed energy density the entropy densities in the cases $x_0=100$ and $x_0=0.032$ are the same. Therefore, the case $x_0=0.032$ may be considered as a counterpart of the case $x_0=100$ with inversed role of pressures.

In all considered cases we assume that the initial energy density in the transverse plane, $\varepsilon_0(\tau_0,{\bf x}_\perp)$, is proportional to the normalized density of sources $\tilde{\rho}(b,{\bf x}_\perp)$, where $b$ is the impact parameter corresponding to a given centrality class, 
\begin{equation}
\varepsilon_0(\tau_0,{\bf x}_\perp) = \varepsilon_{\rm i} \, \tilde{\rho}(b,{\bf x}_\perp).
\label{ei1}
\end{equation}
The parameter $\varepsilon_{\rm i}$ is the initial energy density at the center of matter in most central collisions. Its value may be estimated from the standard hydrodynamic calculations \cite{Florkowski:2010zz}. We use the value \mbox{$\varepsilon_{\rm i} = 86.76$ GeV/fm$^3$}. If the considered matter was in equilibrium, its temperature would be equal to 485 MeV. The normalized density of sources $\tilde{\rho}(b,{\bf x}_\perp)$, is constructed as a combination of the wounded-nucleon density $\rho_W(b,{\bf x}_\perp)$ and the density of binary collisions $\rho_B(b,{\bf x}_\perp)$ \cite{Kharzeev:2000ph},
\begin{equation}
\rho(b,{\bf x}_\perp) =  \frac{1-\alpha}{2}\rho_W \left(b,{\bf x}_\perp \right) 
+ \alpha \rho_B \left(b,{\bf x}_\perp \right),
\label{mix}
\end{equation}
\begin{equation}
 \tilde{\rho}(b,{\bf x}_\perp) = \frac{\rho(b,{\bf x}_\perp)}{\rho(0,0)}.
\end{equation}
The distributions $\rho_W \left(b,{\bf x}_\perp \right)$ and $\rho_B \left(b,{\bf x}_\perp \right)$ are calculated for a given centrality class from the Glauber model in the optical approximation. Following the PHOBOS studies of the centrality dependence of the hadron production \cite{Back:2004dy} we take $\alpha=0.14$. In the limit $\alpha \to 0$ our assumption (\ref{mix}) is reduced to the wounded nucleon model \cite{Bialas:1976ed}.

By fixing both the initial energy density and the initial anisotropy parameter, we determine the initial entropy density profile from Eq. (\ref{epsilon2b}), namely
\begin{equation}
 \sigma(\tau_0,{\bf x}_\perp) = 
\varepsilon_{\rm gqp}^{-1} 
\left[ \frac{\varepsilon_{\rm i} \, \tilde{\rho}(b,{\bf x}_\perp)}{r(x_0)} \right].
\label{sig1}
\end{equation}
Here $\varepsilon_{\rm gqp}^{-1}(\varepsilon)$ is the inverse function to $\varepsilon_{\rm gqp}(\sigma)$. The $b$-dependence displayed on the right-hand-side of Eq. (\ref{sig1}) induces centrality dependence of the initial entropy density profiles.

\section{Results}

\subsection{Central collisions}
\label{centr}

\begin{figure}[t]
\begin{center}
\subfigure{\includegraphics[angle=0,width=0.75\textwidth]{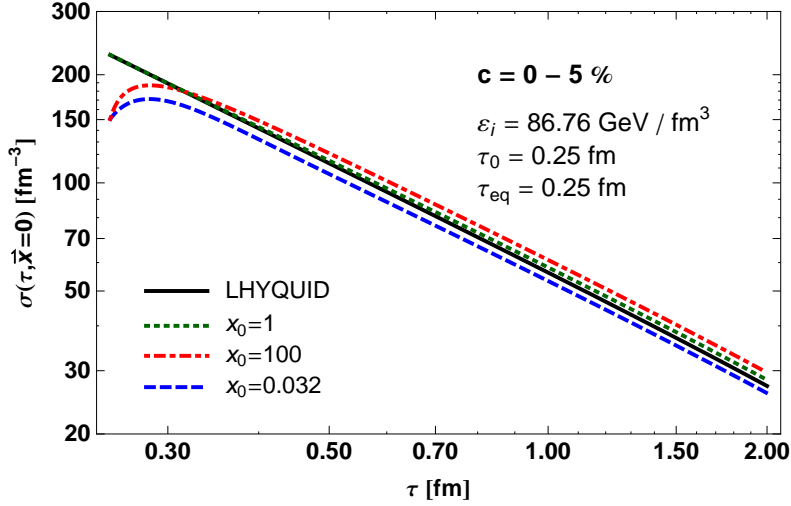}} \\
\end{center}
\caption{\small Time dependence of the entropy density at the center of the system, $\sigma(\tau,{\bf x}_\perp=0)$,  for three different initial values of the anisotropy parameter: $x_0=100$ (dashed-dotted line), $x_0=1$ (dotted line), and $x_0=0.032$ (dashed line). The results have been obtained for the centrality class $c=0-5$\%. The corresponding result of the perfect-fluid hydrodynamics is denoted by the solid line.}
\label{fig:s}
\end{figure}

Let us first discuss our results obtained for the centrality class $c=0-5$\%, which corresponds to the impact parameter $b=2.26$ fm. In our all calculations we use the value \mbox{$\varepsilon_{\rm i} = 86.76$ GeV/fm$^3$}. This implies that the initial entropy density at the center equals  \mbox{$\sigma_{\rm i} = \sigma(\tau_0,{\bf x}_\perp=0)= 227.6$ fm$^{-3}$} for the case \mbox{$x_0=1$}, and  \mbox{$\sigma_{\rm i} = 151.7$ fm$^{-3}$} for the cases $x_0=100$ and $x_0=0.032$. The same initial energy density, \mbox{$\varepsilon_{\rm i} = 86.76$ GeV/fm$^3$}, is assumed in the perfect-fluid calculation based on the LHYQUID code \cite{Chojnacki:2006tv,Chojnacki:2007rq}, which serves as a reference point. Since the perfect-fluid case is obtained formally by setting $x \equiv 0$ in ADHYDRO, the initial entropy density in the perfect-fluid calculation equals \mbox{$\sigma_{\rm i} = 151.7$ fm$^{-3}$} and the initial temperature is \mbox{$T_{\rm i} = 485$ MeV}.

Figure \ref{fig:s} shows the time dependence of the entropy density at the center of the system, i.e., the function $\sigma(\tau,{\bf x}_\perp=0)$, in the time interval \mbox{0.25 fm $\leq \tau \leq$ 2 fm}, for three different initial values of the anisotropy parameter: $x_0=100$ (dashed-dotted line), $x_0=1$ (dotted line), and $x_0=0.032$ (dashed line). The entropy density increases significantly at the initial stage of collisions in the two cases exhibiting strong initial asymmetry, i.e., for $x_0=100$ and $x_0=0.032$. After reaching the maximum at $\tau =$ 0.3 fm,  $\sigma(\tau,{\bf x}_\perp=0)$ starts to decrease. For \mbox{$\tau > $ 0.5 fm}, the entropy density scales approximately as $1/\tau$. This behavior reflects the approximate perfect-fluid behavior reached for significantly large times~\footnote{The scaling $\sigma \propto 1/\tau$ is strictly valid for the longitudinal, boost-invariant expansion. The transverse expansion induces corrections to this behavior of about 10\%.}. In the case $x_0=1$, the entropy density decreases in the very much similar way as in the perfect-fluid case denoted by the solid line. Interestingly, in all considered cases the final entropy densities are very much similar. This result suggests that the final particle multiplicities may be practically the same.

\begin{figure}[t]
\begin{center}
\subfigure{\includegraphics[angle=0,width=0.75\textwidth]{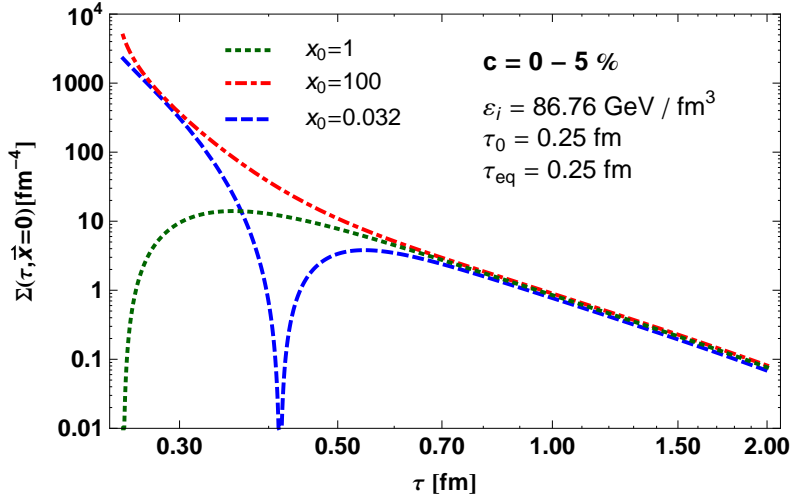}} \\
\end{center}
\caption{\small Time dependence of the entropy production at the center of the system, $\Sigma(\tau,{\bf x}_\perp=0))$, for three different initial values of the anisotropy parameter, notation the same as in Fig. \ref{fig:s} ($c=0-5$\%). }
\label{fig:bigs}
\end{figure}

\begin{figure}[t]
\begin{center}
\subfigure{\includegraphics[angle=0,width=0.75\textwidth]{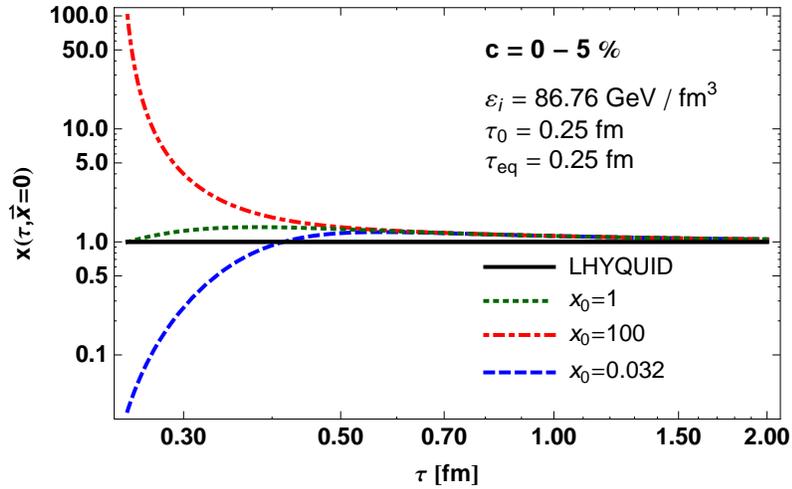}} 
\end{center}
\caption{\small Time dependence of the anisotropy parameter at the center of the system, $x(\tau,{\bf x}_\perp=0)$, for three different values of $x_0$, notation the same as in Fig. \ref{fig:s} ($c=0-5$\%).}
\label{fig:x}
\end{figure}

The behavior of the entropy density may be discussed from a complementary point of view if we analyze the function  $\Sigma(\tau,{\bf x}_\perp=0)$. The time dependence of the entropy source is shown in Fig. \ref{fig:bigs}. The notation is the same as in Fig. \ref{fig:s}. In the case $x_0=100$ (dashed-dotted line) the initial entropy production is large. At later times $\Sigma(\tau,{\bf x}_\perp=0)$ becomes smaller, since the pressure asymmetry decreases; see Fig. \ref{fig:x} where the corresponding time dependence of $x(\tau,{\bf x}_\perp=0)$ has been shown. In the case $x_0=0.032$ (dashed line) the initial entropy production is also large but it drops to zero at the time when $x$ passes unity. Later it increases and becomes similar to that found for the case $x_0=100$. If the initial asymmetry is 1 (dotted line), the entropy production is generally small because $x$ stays close to unity during the whole evolution.

We stress that the results obtained with ADHYDRO where $x$ has been initially set equal to zero are slightly different from the results obtained within perfect-fluid hydrodynamics. The origin of this difference may be traced back to the terms in the evolution equation for $x$, which describe free streaming. In the 1+1 boost-invariant case with Eqs. (\ref{epsilon2a})--(\ref{PL2a}), which is easy to analyze explicitly \cite{Florkowski:2010cf}, the time dependence of $x$ is determined by the equation (\ref{x3}). 

The first term on the right-hand-side of (\ref{x3}) describes the effects of free streaming, while the second term is responsible for thermalization/isotropization. For small initial values of $x$, the first term dominates the very early dynamics and $x$ starts to grow with time. However, as the evolution time increases, the first term becomes negligible, the second term takes over and determines an approach to equilibrium. For large initial values of $x$ the second term dominates the dynamics of matter from the very beginning. If we assume $x=1$ at the beginning of the evolution, the second term vanishes, since $H(1)=0$, and $x$ increases. This effect is important only if $\tau$ is small, which agrees with the idea that the free streaming of partons may be relevant only at the beginning of the evolution of matter, when the system is very much dense and the effects connected with asymptotic freedom may play a role.  

\begin{figure}[t]
\begin{center}
\subfigure{\includegraphics[angle=0,width=0.65\textwidth]{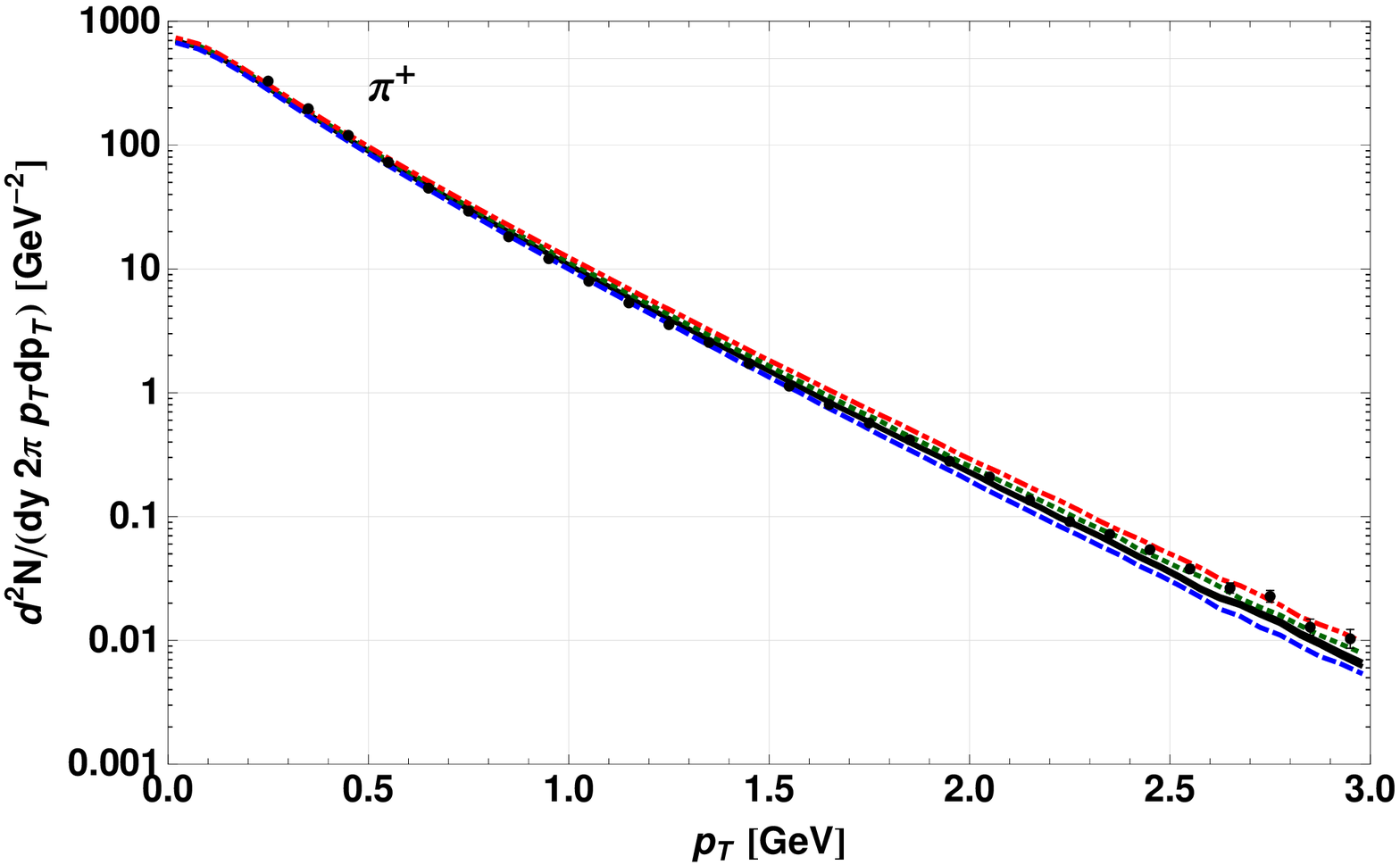}} \\
\subfigure{\includegraphics[angle=0,width=0.65\textwidth]{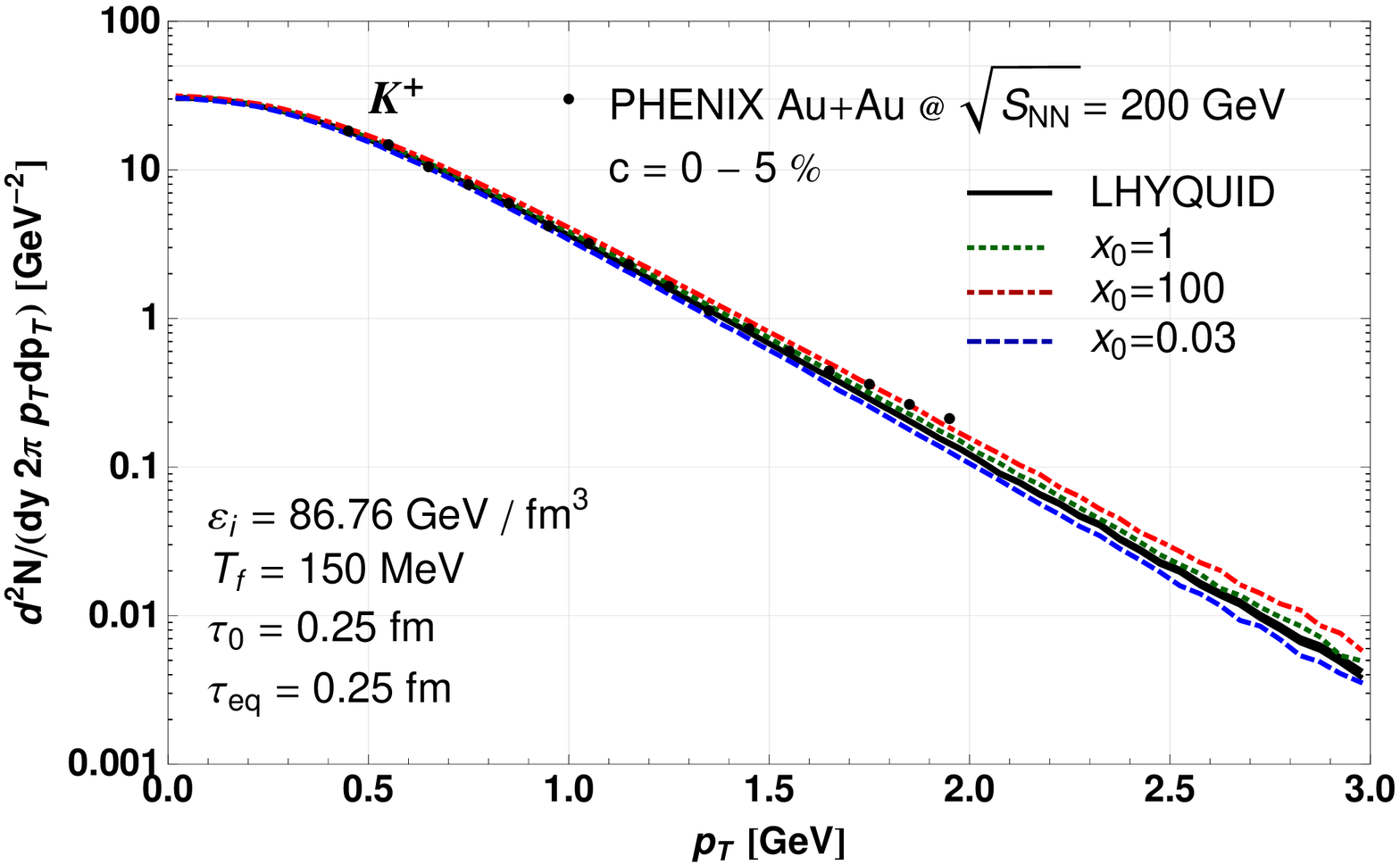}} \\
\subfigure{\includegraphics[angle=0,width=0.65\textwidth]{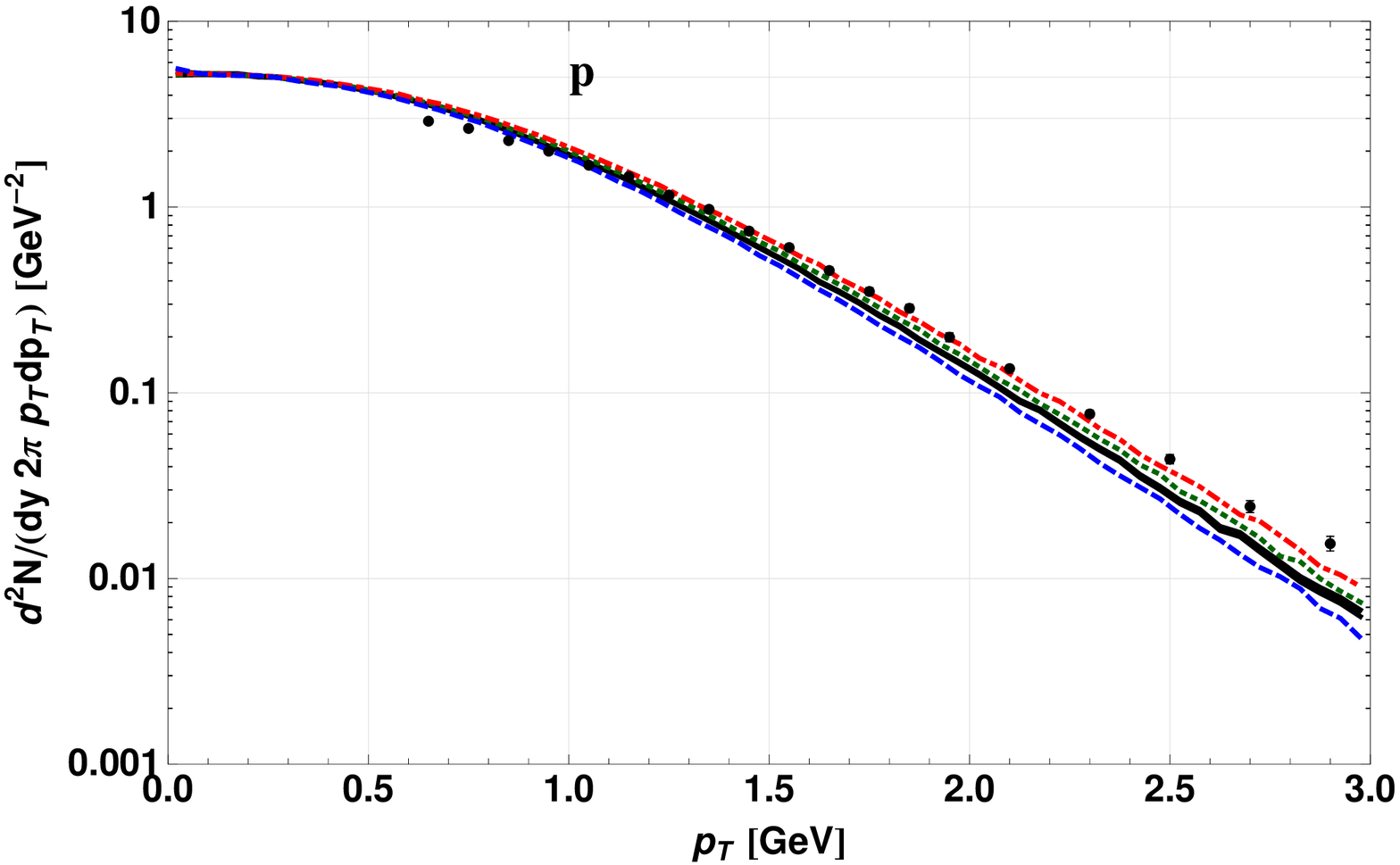}} 
\end{center}
\caption{\small Transverse-momentum spectra of pions (upper part), kaons (middle part), and protons (lower part) for Au+Au collisions at $\sqrt{s_{\rm NN}}=200$ GeV and the centrality class $c=0-5$\%. Notation the same as in the previous Figures. The data are taken from \cite{Adler:2003cb}.}
\label{fig:spectra0005}
\end{figure}

Having solved numerically the ADHYDRO equations for given initial conditions we determine the freeze-out hypersurfaces. We follow the same procedure as in standard 2+1 hydrodynamics. The only difference is that we exclude times smaller than 1 fm, where the system is substantially anisotropic and consists of gluons rather than hadrons that can be freely emitted. For $\tau > 1$ fm, the matter is practically in local equilibrium and we may specify the freeze-out condition by fixing the freeze-out temperature $T_f$. In our calculations we assume $T_f=150$ MeV, the value which turned out be successful in the description of the RHIC data in our earlier approaches. If the freeze-out hypersurface is determined, we use it as an input in  {\tt THERMINATOR 2} \cite{Kisiel:2005hn,Chojnacki:2011hb} to generate physical events.  

Since the analysis of hadronic abundances indicates that the baryon number density is not exactly zero at midrapidity in Au+Au collisions at $\sqrt{s_{\rm NN}}=200$ GeV, we include small non-zero values of the chemical potentials \cite{Florkowski:2001fp}. 

In Fig.~\ref{fig:spectra0005} we show the transverse-momentum spectra of pions (upper part), kaons (middle part), and protons (lower part). We use the same notation as in the previous Figures: $x_0=100$ (dashed-dotted line), $x_0=1$ (dotted line), $x_0=0.032$ (dashed line), LHYQUID (solid line). One can see that all versions of the calculations yield very similar results for the spectra. A small persistent difference can be noticed ---  the results with $x_0 = 100$ are the highest, while those with $x_0=0.032$ are the lowest. In particular, the spectra obtained with $x_0 = 100$ agree better with the data in the high $p_\perp$ region. Nevertheless, the results obtained with different values of $x_0$ are close to each other and to the standard 2+1 hydrodynamic result obtained with LHYQUID. This means that the strong early anisotropic behavior have small impact on the observables. This, in turn, leads us to the conclusion that the data does not exclude a possibility that such an anisotropic stage can exist for about 1 fm at the beginning of the collision. It is also preferable, that the initial transverse pressure is larger than the longitudinal pressure. 

\begin{figure}[t]
\begin{center}
\subfigure{\includegraphics[angle=0,width=0.65\textwidth]{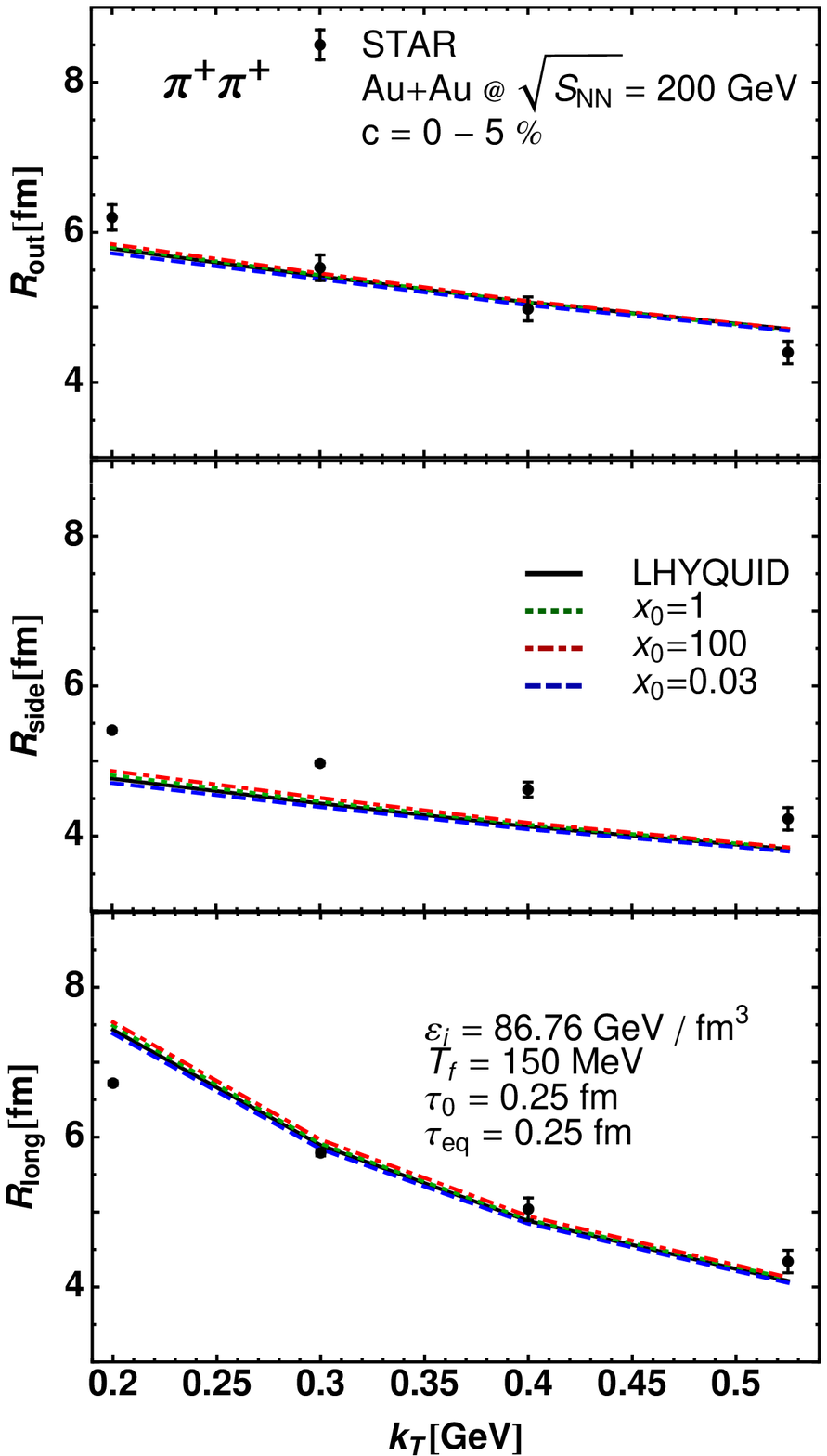}} 
\end{center}
\caption{\small HBT radii calculated with ADHYDRO and LHYQUID. The model results for Au+Au collisions at $\sqrt{s_{\rm NN}}=200$ GeV and the centrality class $c=0-5$\% are compared to the RHIC experimental data \cite{Adams:2004yc}. }
\label{fig:Rcomboc0005}
\end{figure}

\begin{figure}[t]
\begin{center}
\subfigure{\includegraphics[angle=0,width=0.65\textwidth]{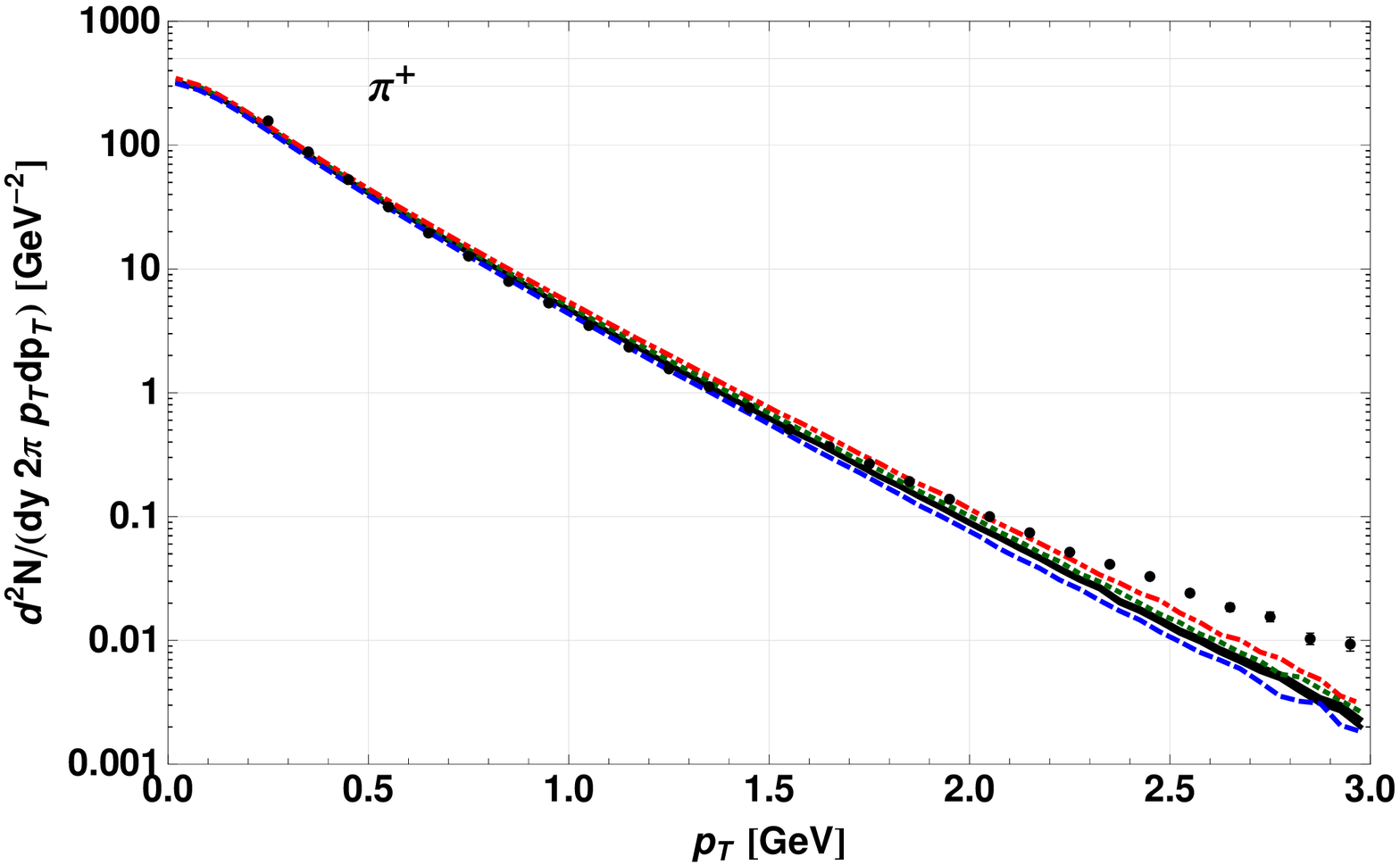}} \\
\subfigure{\includegraphics[angle=0,width=0.65\textwidth]{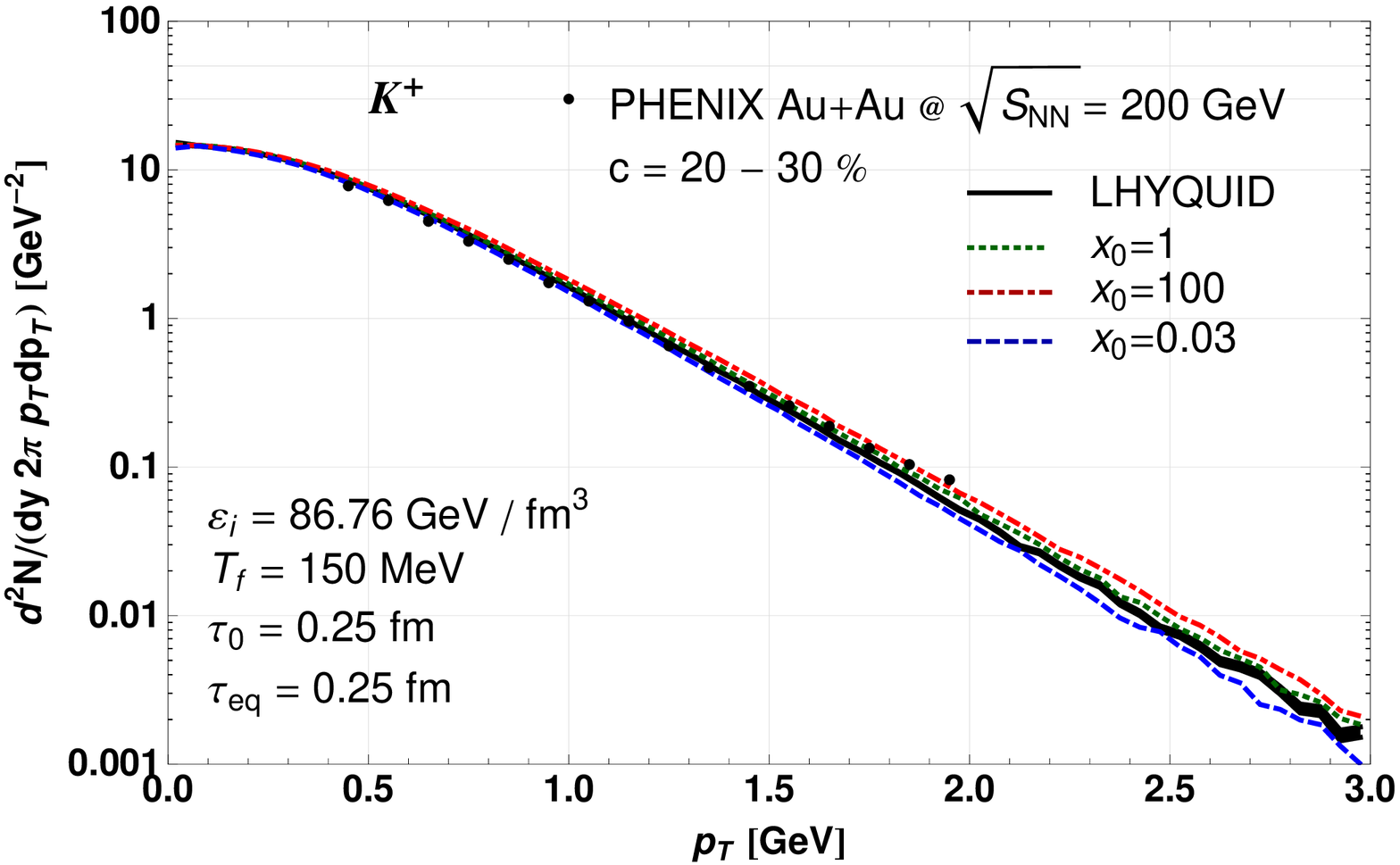}} \\
\subfigure{\includegraphics[angle=0,width=0.65\textwidth]{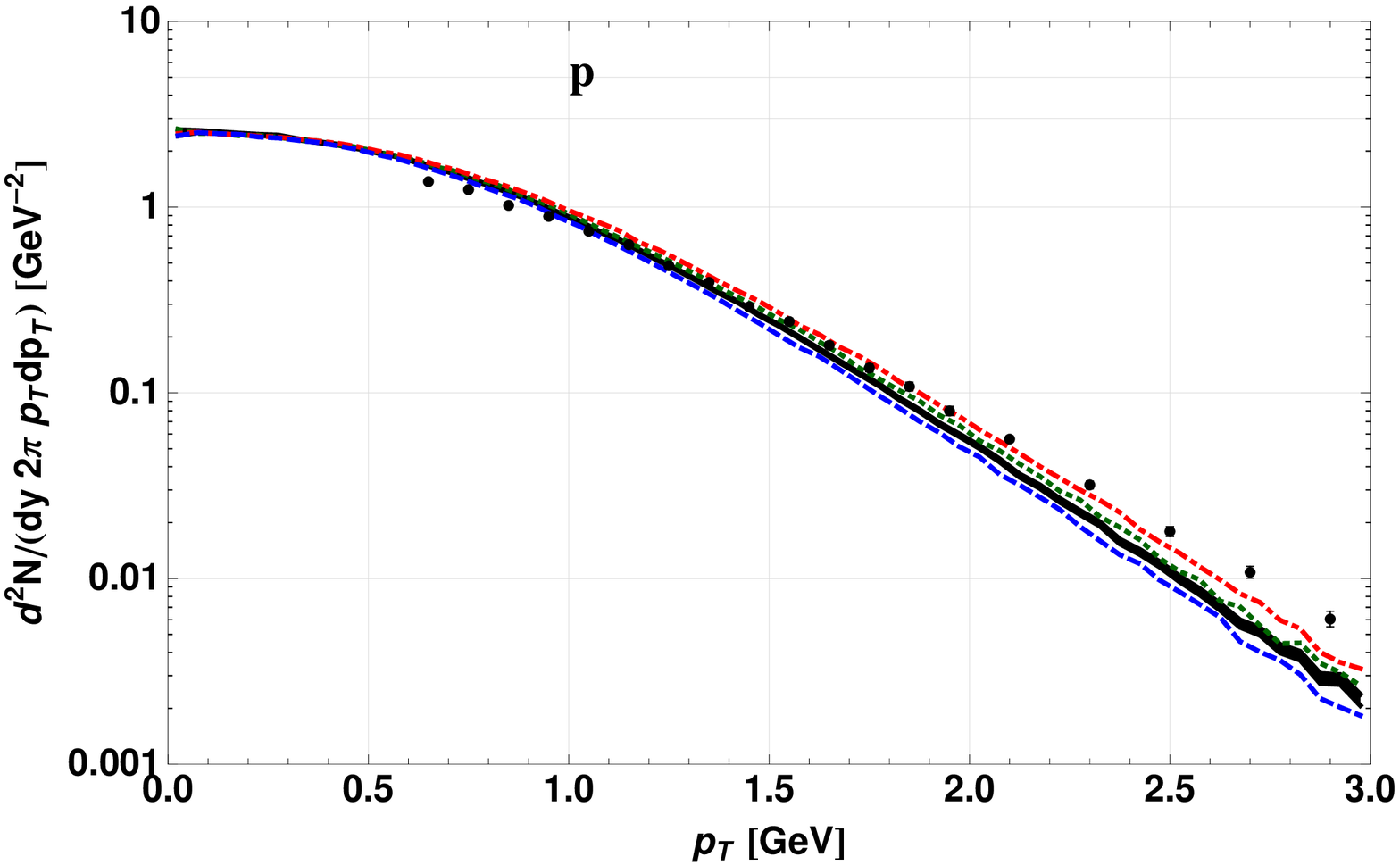}} 
\end{center}
\caption{\small Transverse-momentum spectra of pions (upper part), kaons (middle part), and protons (lower part) for Au+Au collisions at $\sqrt{s_{\rm NN}}=200$ GeV and the centrality class $c=20-30$\%. Notation the same as in previous Figures. The data are taken from \cite{Adler:2003cb}. }
\label{fig:spectra2030}
\end{figure}

Figure \ref{fig:Rcomboc0005} shows the HBT radii calculated with FEMTO THERMINATOR, a part of the THERMINATOR 2 package \cite{Chojnacki:2011hb}. FEMTO THERMINATOR implements the methods described in more detail in Ref. \cite{Kisiel:2006is}. We observe again a small dependence of the model results on the initial value of $x$. The agreement between the model results and the data is very good for $R_{\rm out}$ and $R_{\rm long}$. On the other hand, the model results for $R_{\rm side}$ are about 10\% below the experimental values. Clearly, the HBT puzzle is not completely eliminated in our approach. The presence of the initial, strongly anisotropic stage does not help to improve the agreement with the data that has been already obtained with the LHYQUID calculation. Nevertheless, the overall agreement with the model results and the data is quite satisfactory. 

\subsection{Non-central collisions}

In this Section we analyze non-central collisions. Two centrality classes are considered: $c=20-30$\% and $c=20-40$\%, which correspond to the impact parameter $b=7.16$ fm and $b=7.84$ fm, respectively. With the value \mbox{$\varepsilon_{\rm i} = 86.76$ GeV/fm$^3$} used in Eq. (\ref{sig1}), the $b$-dependence of the source density implies the following values of the central initial entropy density: \mbox{$\sigma_{\rm i} = 183.1$ fm$^{-3}$} for the case $x_0=1$, and  \mbox{$\sigma_{\rm i} = 120.3$ fm$^{-3}$} for the cases $x_0=100$ and $x_0=0.032$ if $c=20-30$\%; \mbox{$\sigma_{\rm i} = 171.3$ fm$^{-3}$} for the case $x_0=1$, and  \mbox{$\sigma_{\rm i} = 113.8$ fm$^{-3}$} for the cases $x_0=100$ and $x_0=0.032$ if $c=20-40$\%.

\begin{figure}[t]
\begin{center}
\subfigure{\includegraphics[angle=0,width=0.65\textwidth]{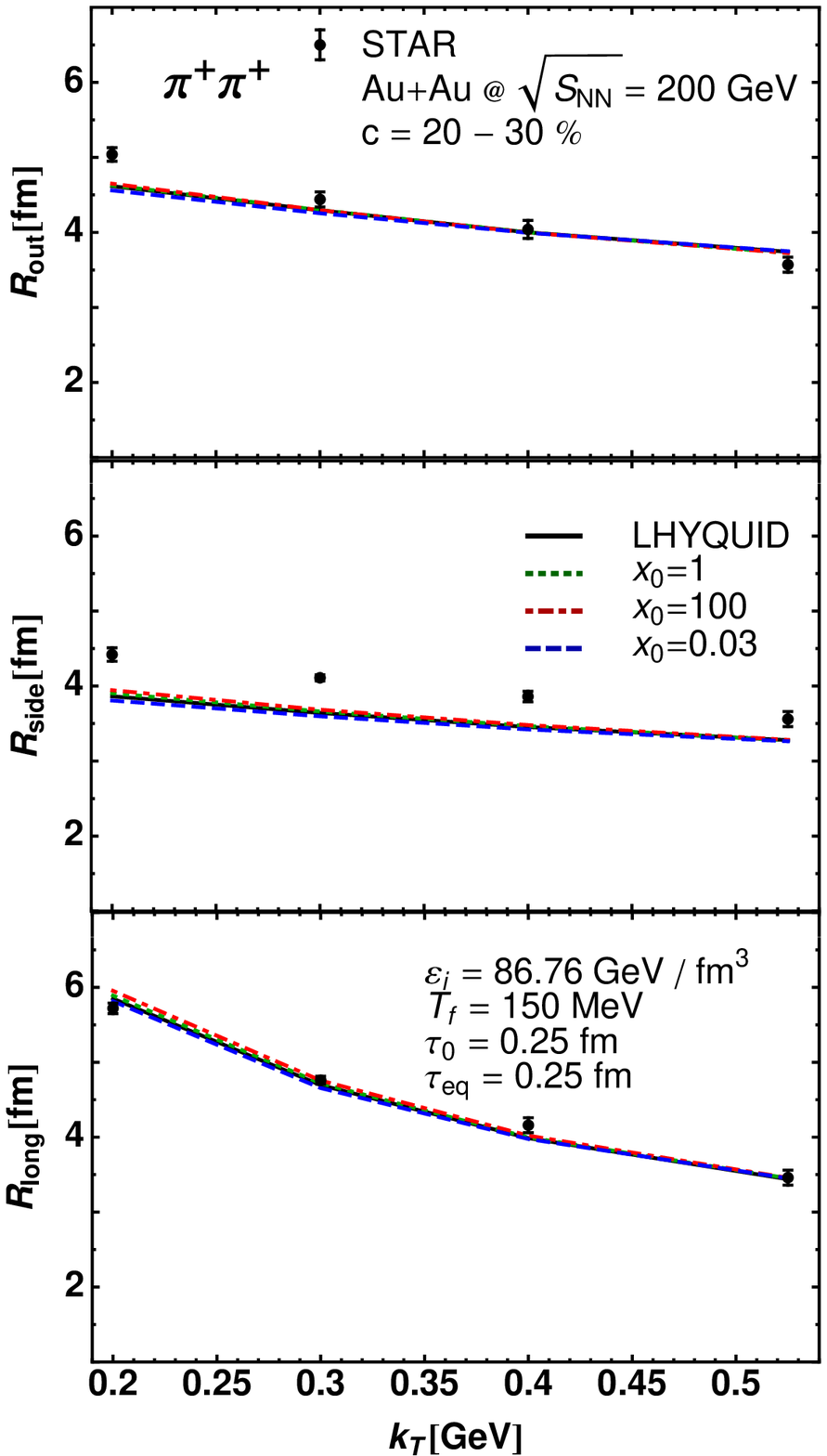}} 
\end{center}
\caption{\small HBT radii calculated with ADHYDRO and LHYQUID. The model results for Au+Au collisions at $\sqrt{s_{\rm NN}}=200$ GeV and the centrality class $c=20-30$\% are compared to the RHIC experimental data \cite{Adams:2004yc}. }
\label{fig:Rcomboc2030}
\end{figure}

\begin{figure}[t]
\begin{center}
\subfigure{\includegraphics[angle=0,width=0.75\textwidth]{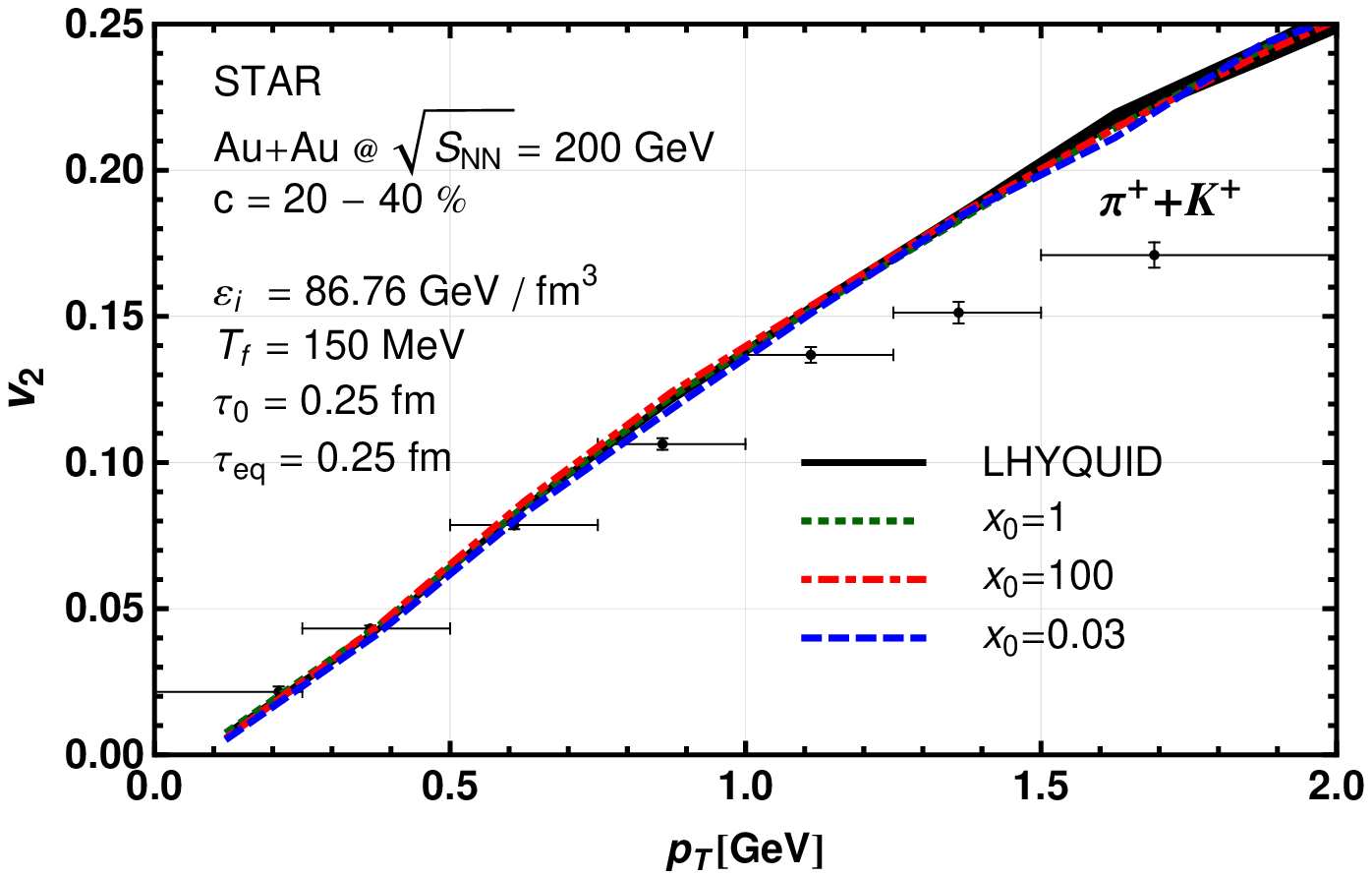}} 
\end{center}
\caption{\small  The elliptic flow of pions and kaons in Au+Au collisions at $\sqrt{s_{\rm NN}}=200$ GeV and the centrality class $c=20-40$\%. The model results  are compared to the RHIC experimental data \cite{Adler:2003kt}. }
\label{fig:pi_v2}
\end{figure}

In Fig. \ref{fig:spectra2030} we show our model results for the transverse-momentum spectra of pions (upper part), kaons (middle part), and protons (lower part), obtained for the centrality class $c=20-30$\% and compared with the RHIC data. One can notice the overall good agreement between the model results and the data. Interestingly, the scaling of the initial entropy based on Eq. (\ref{sig1}) ensures the good normalization of the spectra. Similarly to the central collisions, the case with $x_0=100$ describes slightly better the high $p_\perp$ tails, although the differences between the model and the data in this region are larger than in central collisions. In any case, for \mbox{$p_\perp \leq 2$ GeV}, where the hydrodynamics model is expected to work, the agreement with the data is very good.

Figure \ref{fig:Rcomboc2030} shows the HBT radii for the centrality class $c=20-30$\%. The agreement with the data is similar as in the case of the central collisions. Again, no significant effects of the early anisotropic phase are seen. 

Finally, in Fig. \ref{fig:pi_v2} we show the differential elliptic flow $v_2(p_\perp)$ of pions and kaons for the centrality class $c=20-40$\%. The results obtained with different initial values of the anisotropy parameter coincide with each other. They are also consistent with the LHYQUID result. For \mbox{$p_\perp \leq 2$ GeV} the model results agree with the data. For larger values of transverse-momentum, the perfect-fluid hydrodynamics overshoots the experimental data, which is a known effect. 

\section{Conclusions}

In this paper, a recently formulated model of highly-anisotropic and strongly-dissipative hydrodynamics (ADHYDRO) has been used to study the effects of strong initial anisotropy of pressure on soft hadronic observables studied at RHIC (transverse-momentum spectra, the elliptic flow coefficient $v_2$, and the HBT radii). We have found that the initial conditions with different anisotropies lead to similar results, provided the initial energy density profile in the transverse plane is the same. This result agrees well with earlier findings where the perfect-fluid stage starting at about 1 fm was preceded by free streaming of partons or by expansion of matter thermalized only in the transverse direction. The main conclusion from those studies is that the complete thermalization of matter may take part at the times of about 1 fm. In order to reproduce the elliptic flow of pions, it is not necessary to assume that matter thermalizes within a fraction of one fermi, as the decrease of spatial eccentricity of the reaction zone is compensated by the building of the transverse flow in an early non-equilibrium stage (described as free streaming, transverse hydrodynamics, or highly-anisotropic and strongly-dissipative hydrodynamics used in this paper).

\section{Appendix: Integral form of the conservation laws}

The conservation laws (\ref{enmomcon}) and (\ref{engrow}) can be integrated over the spatial transverse coordinates. This leads to the global conservation laws. In the case of Eq. (\ref{engrow}) we obtain
\begin{equation}
\frac{\partial}{\partial \tau} \int d^2 x_\perp 
\left[ \sigma ( \tau,{\bf x}_\perp ) u^0( \tau,{\bf x}_\perp ) \tau  \right] =
\tau \int d^2 x_\perp \Sigma( \tau,{\bf x}_\perp ).  
\label{g1}
\end{equation}
If the entropy source term $\Sigma( \tau,{\bf x}_\perp )$ vanishes, the left-hand-side of Eq. (\ref{g1}) is zero. This implies that the entropy per unit rapidity is conserved.  In the case of Eq. (\ref{engrow}) we define the quantity $\varepsilon_\perp$, 
\begin{eqnarray}
\varepsilon_\perp( \tau,{\bf x}_\perp ) &=& \varepsilon ( \tau,{\bf x}_\perp ) u_0^2( \tau,{\bf x}_\perp ) +P_\perp( \tau,{\bf x}_\perp )
u_\perp^2( \tau,{\bf x}_\perp ). 
\end{eqnarray}
The integration of (\ref{enmomcon}) yields
\begin{equation}
\frac{\partial}{\partial \tau} \int d^2 x_\perp  \left[ \varepsilon_\perp( \tau,{\bf x}_\perp ) \tau \right]
= -  \int d^2 x_\perp  P_\parallel ( \tau,{\bf x}_\perp ).
\label{g2}
\end{equation}
If the longitudinal pressure $P_\parallel ( \tau,{\bf x}_\perp )$ vanishes, no work is done in the longitudinal direction, and Eq. (\ref{g2}) leads to the conservation of the energy per unit rapidity. Equations (\ref{g1}) and (\ref{g2}) are useful for checking the numerical code. We have found that in our calculations they are satisfied with the accuracy of about 0.001\%.


\end{document}